\documentstyle[11pt,appb,epsf]{article} 

\begin{document}
\title{NON-PERTURBATIVE VEVs FROM A LOCAL EXPANSION
}
\author{A. Jaramillo$^a$ and P. Mansfield$^b$
\address{$^a$Depto. de F\'{\i}sica Teorica--IFIC, Universidad de Valencia.\\ 
Dr. Moliner 50. 46100-Burjassot, Spain.\\
$^b$Dept. of Mathematical Sciences, University of Durham.\\ 
South Road, Durham, DH1 3LE, England.
}
}
\maketitle
\begin{abstract}
We propose a method for the calculation of  vacuum expectation values (VEVs) 
given a non-trivial, long-distance vacuum wave functional (VWF) of the kind 
that arises, for example, in variational calculations. The VEV is written in 
terms of a Schr\"odinger-picture path integral, then a local expansion for 
(the logarithm of the) VWF is used. The integral is regulated with an explicit
momentum cut-off, $\Lambda$. The resulting series is not expected to converge 
for $\Lambda$ larger than the mass-gap but studying the domain of analyticity
of the VEVs allows us to use analytic continuation to estimate the 
large-$\Lambda$ limit. Scalar theory in $1+1$ dimensions is analyzed, where 
(as in the case of Yang-Mills ) we do not expect boundary divergences.
\end{abstract}
  
\section{Introduction}

The use of the Schr\"odinger picture in field theory provides a natural
framework for non-perturbative calculations based on ans\"atze for the 
vacuum wave functional (VWF). Although the Yang-Mills VWF has been  analyzed
numerically by lattice simulations \cite{greensite-lat,arisue}
and analytical approximations \cite{greensite79,paul,nair}, it is an open
problem to compute VEVs with such a VWF because  we have to address
 the calculation of non-Gaussian path integrals. One way to 
systematically compute the path integral
\begin{equation}
\int {\cal D}\phi \,\, A[\phi]\, |\Psi_0[\phi]|^2
\label{vev}
\end{equation}
would be to expand $\Psi_0[\phi]$ about a Gaussian, so we can use 
Wick's theorem to obtain the VEV. In principle, this would involve a 
perturbation expansion with (dressed) propagators and an infinite number 
of non-local vertices with increasing dependence on high momenta. Within 
perturbation theory, it can be explicitly shown that these diagrams will not 
generate new divergences other than the ones which are substracted by 
renormalization \cite{paper}. Outside perturbation theory we may encounter 
divergences (which will have to cancel when we resum the expansion) due
to the fact that the Gaussian term (around which we expand) does not damp
 the high momentum modes strongly enough. We will show how to compute such
a path integral by performing a further expansion in terms of local expressions. Therefore, 
we will give a method to compute finite VEVs, from a given $\Psi_0[\phi]$,
by expanding the logarithm of $\Psi_0[\phi]$ in power of $\phi$ and its
derivatives (and considering the quadratic term in $\phi$ as the unperturbed
part). Notice that now (if we compute an equal-time VEV) the $|\Psi_0[\phi]|^2$
can be interpreted as the exponential of (minus) a euclidean local action
(which lives in the quantization surface $t=0$) with an infinite number of 
(non-renormalizable) terms which apparently generate ultra-violet divergences. 
These arise because the path integral includes configurations with momentum
far beyond the convergence
radius of our  local expansion.
In \cite{paper} we have argued that
that (provided the theory has a mass-gap) the convergence radius
of the local expansion is non-zero and finite. Therefore, we introduce an
explicit cut-off, $\Lambda$, on the Fourier components of the field 
configurations of Eq. (\ref{vev}). Now the equal-time VEV of an operator 
can be computed from the $\Psi_0$-local expansion, provided $\Lambda$ is
smaller than the convergence radius. Of course, we want to send $\Lambda$
to infinity. In \cite{paper} we have studied the domain of analyticity  of the
(equal-time) VEVs as functions of this cut-off and used it to compute the 
$\Lambda\rightarrow\infty$ value by analytic continuation. The particular
method chosen for the continuation is not essential, although the procedure
should be suitable for numerical analysis. We found to be most convenient a method
related to Borel resummation (which itself is commonly used to re-sum asymptotic series).
We will discuss the method in the context of scalar theory in $1+1$
dimensions and we will give an example based on 
perturbation theory (where we can check it).

\section{Analyticity in the Schr\"odinger cut-off}
In \cite{paul2} it was shown that for Yang-Mills theory $\Psi_0[A(x/\sqrt{s})/\sqrt{s}]$, 
is analytic in the cut $s$-plane 
with the cut in the negative real axis. The same holds for 
scalar theory \cite{paper}. This analyticity of the scaled VWF provides
a method to recover the full $\Psi_0[\phi]$ from its functional Taylor
expansion around slowly varying configurations. This analyticity can be
extended to scaled equal-time VEVs \cite{paper}, so they can be reconstructed
from the resummation of a perturbation expansion (the expansion parameter will
be $\Lambda/m$, with $m$ the mass scale which appears in the chosen VWF and
which will be related to the energy spectrum)generated by locally expanding 
the VWF. 

\begin{figure}[t]
\begin{center}
\leavevmode
\epsfxsize=5.0cm
\epsfysize=5.0cm
\epsfverbosetrue
\epsffile{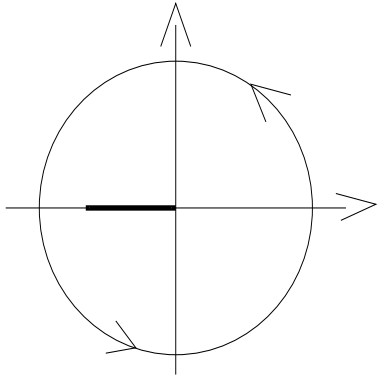}
\end{center}
\caption{ Large radius contour for the integral $I(\lambda)$. }
\label{fig1}
\end{figure}

For scalar theory in $1+1$ dimensions, we have shown \cite{paper}
that
\begin{equation}
K(s)=s^{n/2} \langle\Psi_0|A[\tilde\phi(\frac{p}{\sqrt{s}})]|\Psi_0
\rangle_{\frac{\Lambda}{\sqrt{s}}}
\end{equation}
is analytic in the cut $s$-plane, with the cut going from
$s=-\Lambda^2$ to $s=0$ (and with $n$ being a integer number).
We will recover $K(1)$ (which gives an approximation to the VEV if
$\Lambda$ is large enough) by analytic continuation from $s=\infty$
(where a local expansion can be used) to $s=1$. The method for the
analytic continuation is based on constructing a  function, $I(\lambda)$,
from $K(s)$:
\begin{equation}
I(\lambda)=\frac{1}{2\pi i}\int_C ds\,\, \frac{e^{\lambda(s-1)}}{s-1}
\, \tilde K(\frac{p_i}{\sqrt{s}},\frac{\Lambda}{\sqrt{s}}).
\label{Ilambda}
\end{equation}
The integration contour $C$ is shown in Fig.~\ref{fig1} together with
the $[-\Lambda^2,0]$ cut.
A local expansion of the VWF will give a
series in $p^k$
\begin{equation}
\tilde K=\sum_k a_k
\frac{1}{\sqrt{s}^k}=\sum_k b_k \frac{1}{(s-1)^k},
\end{equation}
which has been rearranged in a series in $1/(s-1)^n$, so the
Eq.(\ref{Ilambda}) will give
\begin{equation}
I(\lambda)=\sum_n b_n \frac{\lambda^n}{n!}.
\label{Borel}
\end{equation}
It is shown in \cite{paper} (by collapsing the contour to an
infinitesimal circle around $s=1$ and a contribution from the cut,
which will be damped by the exponential term) that the
$\lambda\rightarrow\infty$ limit of Eq.(\ref{Ilambda})  gives the VEV:
\begin{equation}
\tilde K(p_i,\Lambda)=\lim_{\lambda\rightarrow\infty} I(\lambda)=
\lim_{\lambda\rightarrow\infty} \sum_n b_n  \frac{\lambda^n}{n!}.
\end{equation}
Because we expect that the convergence radius of Eq.(\ref{Borel})
will be infinite \cite{paper}, we can obtain an approximation for
$I(\infty)$ by truncating the (alternating) series at $N$ terms, 
and taking $\lambda$ as large as is consistent with this.
In practice this means that we take  
$b_N \lambda^N/N!$ to be a small fraction of the value of the truncated sum.
We
have to assume that the function $I(\lambda)$ does not have a plateau
at finite $\lambda$, which could incorrectly indicate that the
$I(\infty)$ limit has been reached (and therefore the truncated
approximant would give the value of $I(\lambda)$ at the plateau,
instead of $I(\infty)$). We may instead estimate $I(\infty)$ by looking at the
point where the approximant (a polynomial in $\lambda$) is stationary,
then we have to assume that $I(\lambda)$ has no stationary point other
than the one at $\lambda\rightarrow\infty$.

\section{Example}
\begin{figure}[t]
\begin{center}
\leavevmode
\epsfxsize=10.0cm
\epsfysize=3.0cm
\epsfverbosetrue
\epsffile{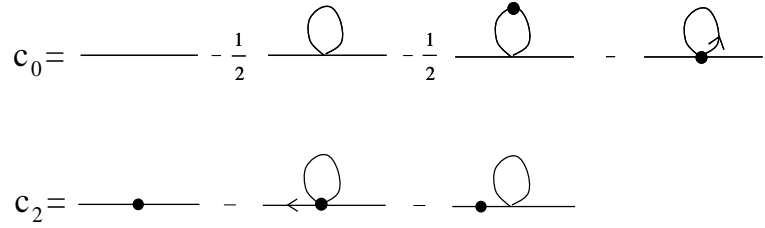}
\end{center}
\caption{ Diagrams for the lowest $\Lambda^n$ terms of $c_0$ and $c_2$.}
\label{fig2}
\end{figure}

In this section we will illustrate how to compute the, equal-time,
two-point function in a $1+1$-dimensional scalar theory model where the
VWF is given, for slowly varying configurations, by
\begin{eqnarray}
|\Psi_0|^2 &=& N
\exp\left(-\frac{1}{2}\int\tilde\phi\tilde\phi(\alpha_0+
\nonumber\right.  \\ &&\left.\alpha_2
p^2+\cdots)-\frac{1}{4!}\int\tilde\phi\tilde\phi\tilde\phi
\tilde\phi(\beta_0+\beta_2\sum_i p_p^2+\cdots)\right).
\label{localVWF}
\end{eqnarray}
And we assume that this VWF has a sensible UV limit (high frequency
modes for the configurations), so the analyticity of the VEV
(which was proven in \cite{paper} for the true vacuum) also would hold
here. We write the (connected) two-point function for small momenta as
\begin{equation}
\langle\Psi_0|\tilde\phi(p)\tilde\phi(-p)|\Psi_0\rangle = c_0+c_2
p^2+\cdots
\end{equation}
the coefficients $c_0$ and $c_2$ can be computed by performing a
perturbation expansion with a small cut-off $\Lambda$. They will be
given by a series in $\Lambda$. In  Fig.~\ref{fig2} we give the result
of the computation of $c_0$ and $c_2$ until order $O(\Lambda^5)$ and
$O(\Lambda^2 p^2)$ respectively.
Where the propagators are $1/\alpha_0$, the momenta insertion
(denoted by a dot) are $-\alpha_2 q^2$ and the dotted vertex with
arrow is $\beta_2 q^2$ (the arrow shows where the momentum is sitting).
In $\phi^4$ theory at first order in perturbation theory, the VWF has a
good UV limit \cite{paper} and the VEV analyticity holds. The values
for the $\alpha_0$, $\alpha_2$, $\beta_0$ and $\beta_2$ are given by
\begin{eqnarray}
\alpha_0&=&2m \nonumber \\
\alpha_2&=&(m-\frac{m}{12\pi}\frac{g}{m^2})\frac{1}{m^2} \nonumber \\
 \beta_0&=&\frac{m}{2}\frac{g}{m^2}\nonumber \\
\beta_2&=&-\frac{m}{16}\frac{g}{m^2}\frac{1}{m^2}
\end{eqnarray}
and the Fig.~\ref{fig2} gives for $c_0$
\begin{equation}
\frac{1}{2m}-\frac{1}{32\pi m}\frac{g}{m^2}\frac{\Lambda}{m}+
\frac{1}{128\pi m}\frac{g}{m^2}\frac{\Lambda^3}{m^3}+\frac{1}{m}
O(\frac{g^2}{m^4}\frac{\Lambda^5}{m^5})
\end{equation}
and after resummation
\begin{equation}
c_0=\frac{1}{2m}-\frac{1.88}{32\pi m}\frac{g}{m^2}
\end{equation}
to be compared with
\begin{equation}
c_0=\frac{1}{2m}-\frac{2}{32\pi m}\frac{g}{m^2}
\end{equation}
We need to go to the next order in $\Lambda$ in order to be able to give a
resummed value for $c_2$. Finally,
in \cite{paper} we have shown that $\alpha_0\neq 0$ provided
$\dot\phi(0)|0\rangle\neq 0$.
\section{Conclusion}
We have shown how to compute VEVs by using a
local expansion of the true VWF within the context of scalar theory in $1+1$ dimensions,
where no extra counterterms are needed (this feature is also shared by 
Yang-Mills theory in $3+1$ dimensions).  In \cite{paper} we have shown how the
analyticity properties of the true vacuum in the cut-off, $\Lambda$, can be
used to recover the VEV from a series in positive powers of $\Lambda$ 
(which is the usual
output of a local expansion of the VWF). If we want to use a variational
ansatz for the VWF, we will have to asssume (or prove) that its short distance
behaviour is such that the $\Lambda\rightarrow\infty$ is finite
and, furthermore, that the analyticity behaviour of VEVs in $\Lambda$ is
preserved (so we can use the analytic continuation). We have given a
diagrammatic approach to compute the first terms of a series in $\Lambda$ for
a VEV in the scalar theory, which (after our resummation) will give the
$\Lambda\rightarrow\infty$ value of the VEV. We expect that this method can
be generalized to Yang-Mills theories in $3+1$ dimensions where, in the strong
coupling limit, we expect to have a local VWF and therefore use the
resummation to compute the VEV of a large Wilson loop.

\section*{Acknowledgments}
A. J. acknowledges NATO for a conference grant and support by DGES under 
contract PB95-1096.

\bibliographystyle{unsrt}

\end{document}